# Polarization rotation theory for field-induced-phase transitions in BaTiO$_3$ single crystal

Hui Zhang[*]

School of Materials Science and Engineering, South China University of Technology, Guangzhou 510640, People's Republic of China

Abstract

We have proposed the polarization rotation theory to understand the underlying physics of the large piezoelectric effect in ferroelectric crystals by referring to the coherent rotation model in ferromagnetism. When both the electric field energy and crystalline anisotropy energy are taken into account, the polarization in BaTiO$_3$ crystal can be rotated toward the field direction. The numerical results have indicated the highly anisotropic polarization rotations and the field-induced-phases with different symmetries in the crystal. This theory is helpful for understanding the polarization reversal and the electromechanical effect in ferroelectric materials.

## I. INTRODUCTION

In the past decade, the polarization rotation mechanism [1-5] has been proposed for understanding the underlying physics of large piezoelectric effect in ferroelectric materials. Despite the fact that the polarization rotation in the rhombohedral -tetragonal phase transition is energetically favorable by the first-principle calculations [3,4], a theory that is physically simple and practical is needed to describe the response of polarization to an electric field. In fact, from the history of ferromagnetism [6], the combination of both domain wall motion [7] and the polarization rotation mechanisms [1-5] can be very helpful for understanding the polarization reversal in ferroelectric crystals. Moreover, the electric behaviors of ferroelectric materials under the combination of both the electric field and the stress [8-16] are very similar to the magnetic behaviors of ferromagnetic materials [17-19], while in ferromagnetism those phenomenological theories have been successfully applied to explain the experimental observations [6,20-22]. This raises a question of whether similar theories with reference to ferromagnetism can be introduced to describe ferroelectric materials. Several questions must be addressed at first. (1) The polarization reversal in ferroelectric crystals is always accompanied with the deformation of the crystal cell. That is, the crystal cell must accordingly deform so that the potential of the system can be lowered [23-28]. Small deformation of the crystal can lead to the change in the symmetry of the crystal, and thus the phase transition occurs. As Park $et\ al.$ [24] have pointed out that in $BaTiO_3$ single crystal the field-induced-polarization rotation can lower the symmetry of the crystal, and lead to the existence of pseudomonoclinic phases such as $P_1^2 \neq P_2^2$, $P_3^2=0$ and $P_1^2=P_2^2 \neq P_3^2$ ($P_1$, $P_2$, and $P_3$ being the components of the polarization). The experimental observations [1,2] have confirmed these field-induced-phase transitions in ferroelectric crystals and the changes in the crystal cell. (2) In ferromagnetism, it is assumed that the saturation magnetization of ferromagnetic materials $M_s$ keeps constant, and is dependent on the temperature and the composition of the materials, i.e., $M_s=(M_x^2+M_y^2+M_z^2)^{1/2}$, $M_x$, $M_y$, and $M_z$ are the components of the saturation magnetization $M_s$ along the $x$, $y$, and $z$ directions, respectively [21]. However, for ferroelectric materials, it is commonly accepted [5,30-32] that the electric field or the stress can induce an additional polarization $\Delta P$, and the polarization $P$ is the sum of the saturation polarization $P_s$ and the field-induced-polarization $\Delta P$, namely $P=\Delta P+P_s$. We assume that at a certain temperature $T_0$ below the Curie temperature $T_c$ ($T_0<T_c$), $P=P_s=(P_1^2+P_2^2+P_3^2)^{1/2}$ always holds constant. The reason for this is that the experimental data have confirmed the anisotropic dependence of the polarization reversal on the crystallographic axis [1,10,12]. For example, the polarization values along the polar directions are always larger than those along other non-polar axes. Therefore, the saturation polarization holding constant does not disagree with the experimental data. (3) For ferroelectric crystals, Landau-Devonshire (LD) theory is the most successful one [30-32]. However, the expression for the free energy in LD theory is very complex, and very difficult to give the clear physical meaning for each term. Furthermore, higher-order terms of the free energy are often needed to explain the emerging physical phenomena. For instance, the eighth-order terms must be added to theoretically describe the experimentally newly discovered monoclinic phases [34]. Therefore, the theory based on the simple mathematical expression can be a good choice for ferroelectric crystals. (4) For ferroelastic crystals, their order parameters such as the magnetization or the polarization can couple with the stress, leading to the changes in their components along the stress direction [17,30]. For isotropic magnetostrictive materials, this coupling is dependent on the product of the saturation magnetostriction of materials $\lambda_s$ and the stress $\sigma$ ($3/2\lambda_s\sigma$). For ferroelectric crystals, the experimental observations have exhibited similar experimental phenomena [8-16], but the corresponding explanations lack the theoretical support. Therefore, the mathematically and physically simple expression for the electromechanical coupling in ferroelectric crystals is needed. On addressing the above issues, with reference to the coherent rotation theory in ferromagnetism, we will investigate the underlying physics of the polarization rotation in ferroelectric crystals.

---

[*] Email address: zhope@scut.edu.cn

## II. THEORETICAL MODEL

We assume that the ferroelectric crystal consists of non-interacting single domains, and each domain has the saturation polarization $P_s$. For $T_0<T_c$, the saturation polarization keeps constant, and is dependent on the temperature and the composition of the materials. Within a single domain, the electric dipoles are parallel to each other. Therefore, for all these single domains, this energy (keeping the electric dipoles parallel to each other) $f_0$ holds constant. There is another energy associated with the crystal structure called the crystalline anisotropy energy $f_a$. This crystalline anisotropy energy determines the orientations of the saturation polarization $P_s$ for each single domain, namely, the spontaneous polar directions of the crystal. The last energy term is the field energy $f_e$, depending on the electric field. This field energy tries to overcome the action of the crystalline anisotropy energy and makes the polarization rotate toward the field direction. After the paraelectric to ferroelectric phase transition, the symmetry of the tetragonal BaTiO$_3$ crystal is the subsystem of the cubic symmetry of the cubic parent phase, and then the crystalline anisotropy energy $f_a$ can be written as [6]

$$f_a = K_1\left(\alpha_1^2\alpha_2^2 + \alpha_2^2\alpha_3^2 + \alpha_3^2\alpha_1^2\right) + K_2\left(\alpha_1^2\alpha_2^2\alpha_3^2\right) \tag{1}$$

where $K_1$ and $K_2$ are the first and second order crystalline anisotropy constants, and $\alpha_1$, $\alpha_2$, and $\alpha_3$ are the direction cosines of the saturation polarization.

The field energy $f_e$ can be given by

$$f_e = -EP_s\left(\alpha_1\beta_1 + \alpha_2\beta_2 + \alpha_3\beta_3\right) \tag{2}$$

where $E$ is the applied electric field, and $\beta_1$, $\beta_2$, and $\beta_3$ are the direction cosines of the applied field.

Thus, the total free energy per unit volume in BaTiO$_3$ crystal $f_t$ is

$$f_t = f_0 + f_a + f_e \tag{3}$$

The equilibrium conditions must satisfy

$$\left.\begin{array}{l}\partial f_t/\partial \alpha_1 = 0 \\ \partial f_t/\partial \alpha_2 = 0\end{array}\right\} \tag{4}$$

and

$$\left.\begin{array}{l}\partial f_t^2/\partial \alpha_1^2 > 0 \\ \dfrac{\partial f_t^2}{\partial \alpha_1^2}\dfrac{\partial f_t^2}{\partial \alpha_2^2} - \left(\dfrac{\partial f_t^2}{\partial \alpha_1 \partial \alpha_2}\right)^2 > 0\end{array}\right\} \tag{5}$$

According to the stable conditions given above, we can obtain the numerical solutions for the direction cosines of the polarization under the equilibrium conditions $\alpha_1$, $\alpha_2$, and $\alpha_3$ [35]. We now can calculate the components of the saturation polarization along the field direction $P$

$$P = P_s\left(\alpha_1\beta_1 + \alpha_2\beta_2 + \alpha_3\beta_3\right) \tag{6}$$

Under the action of the field $E$, the electrostriction along the direction defined by

the direction consines ($\beta_1$, $\beta_2$, $\beta_3$) can be given by ( see Appendix A)

$$\lambda(E) = \frac{3}{2}\lambda_{100}\left(\alpha_1^2\beta_1^2 + \alpha_2^2\beta_2^2 + \alpha_3^2\beta_3^2 - \frac{1}{3}\right) + 3\lambda_{111}(\alpha_2\alpha_3\beta_2\beta_3 + \alpha_3\alpha_1\beta_3\beta_1 + \alpha_1\alpha_2\beta_1\beta_2) \tag{7}$$

where $\lambda_{100}$ and $\lambda_{111}$ are the saturation electrostriction along the [100] and [111] axes for a single crystal, respectively, and $\lambda_{100}=2/3(Q_{11}-Q_{12})P_s^2$ and $\lambda_{111}=1/3Q_{44}P_s^2$ ($Q_{11}$, $Q_{12}$, and $Q_{44}$ being the electrostrictive constants).

In the computation, the values for $P_s$ and $K_1$ can be calculated according to LD theory ( see Appendix A), and $K_2=0$. For example, when the field and the stress are both zero, the saturation polarization along the <100> directions $P_s=0.2716$ C/m² can be obtained [31,32]. From the energy difference between the [111] and [100] axes under the stable conditions, $K_1= 1.207\times10^6$ J/m³. $Q_{11}=0.11$ C$^{-2}$·m⁴, $Q_{12}=-0.043$ C$^{-2}$·m⁴, and $Q_{44}=0.059$ C$^{-2}$·m⁴.

## III. RESULTS AND DISCUSSION

FIG. 1.(Color online). Dependence of the components of polarization of single domains (a) and electrostrictive responses (b) on the field applied along the [001], [011], and [111] axes. In (a), the inset shows the dependence of the components of the polarization in the field range of -50~50 kV/cm, and the solid arrows indicate the polarization jumps for 180º domains while the dashed arrows indicate the polarization jumps for 90º domains. In (b), the dashed arrows indicate the polarization orientations for single domains at the thermally depolarized state.

According to the above model, we calculated the electric hysteresis loop and the electrostrictive response in BaTiO₃ single crystal. Figure 1 shows the dependence of the polarization and the electrostrictive responses for single domains on the field $E$ along the [001], [011], and [111] axes. For BaTiO₃ single crystal, the crystalline anisotropy constant $K_1>0$, and the spontaneous polar directions are the six <100> axes [6]. There are six types of single domains, i.e., [100], [$\bar{1}$00], [010], [0$\bar{1}$0], [001] and

[00$\bar{1}$]-oriented domains. When BaTiO$_3$ single crystal is cooled from the temperature above $T_c$ without the poling field, these six types of domains can coexist in the crystal, with the polarizations equally distributed along each of the six <100> directions. Because the polarization directions for these domains are opposite to each other, the total polarization is zero.

It has been found in Fig. 1 that the electric behaviors for these single domains are different. When the field $E$ is applied along the [001] axis, the polarizations for [001] and [00$\bar{1}$]-oriented domains are parallel and antiparallel to the field direction, and the polarizations for [100], [$\bar{1}$00], [010], and [0$\bar{1}$0]-oriented domains are perpendicular to the field direction. At the thermally depolarized state, there are 180º and 90º domains in BaTiO$_3$ single crystal. When the crystal is poled along the [001] axis, for the electric field cycling of +$E_{max}$→-$E_{max}$, the path for the polarization reversal is given by the solid arrows in Fig. 1(a). For large field strength, the polarization is parallel to the field direction. When the field decreases, the polarization cannot be rotated and remains in their own energy minima due to the energy barrier between different energy minima. At the coercive field $E_c$, this energy barrier decreases to 0, and for $E$>$E_c$ the discontinuous rotation for the polarization from the [001] direction into the [00$\bar{1}$] direction occurs, leading to the abrupt polarization jumps in the electric hysteresis loops. For other four types of domains, when the field increases from 0, the polarizations are first rotated toward the field direction, and then at some critical field abruptly into the field direction. The former two types of domains are 180º domians, and their electrostrictive responses $\lambda(E)$ remain unchanged, leading to no change in electrostriction $\Delta\lambda=\lambda(E)-\lambda(0)$. For 90º domains, when the polarizations are rotated into the field direction, large change in electrostriction $\Delta\lambda\approx$1% can be obtained. For the field $E$ applied along the [111] direction, for [100], [010], and [001]-oriented domains the polarizations are at the same angle of 54.7º with respect to the field direction, and the electrosrictive behaviors are also the same. The polarizations can be rotated from the path for [100], [010], and [001]-oriented domains into that for [$\bar{1}$00],

[0$\bar{1}$0], and [00$\bar{1}$]-oriented domains, and these are not the 90º polarization rotation. Before polarizing to the saturation polarization, the electrostrictive responses for these domains are nearly proportional to the field, and $\Delta\lambda$ is 0.17%. For the field $E$ applied along the [011] direction, the polarizations for [001] and [010]-oriented domains are at an angle of 45º with respect to the field direction, and for [100] and [$\bar{1}$00]-oriented domains, the angles are 90º. The electric behaviors for these domains are similar to those for the field along the [001] and [111] axes.

We also can see that the polarization rotations in BaTiO$_3$ crystal are significantly anisotropic. When polarized along the non-polar directions, the near linear dependence of the electrostrictive responses on the field is also clearly exhibited. In our theory, BaTiO$_3$ single crystal is assumed to be perfect and consist of non-interacting single domains with the polarizations aligned along the <100> axes. However, for the imperfect real single crystal, there are the varying deviations of the initial polarizations from the <100> axes. Our numerical results have shown that the remnent polarizatin can decrease even for a very small angle between the field direction and the <100> axes. Therefore, the real BaTiO$_3$ single crystal can be polarized to the saturation polarization only by very large field strength (larger than the crystalline anisotropy field $E_K=2K_1/P_s=8.9$ MV/m) (see the inset in Fig. 1(a)). The curves between the remnent polarization and the saturation polarization can be regarded as the straight lines. This demonstrates the non-uniformity of polarization reversal for single domains with different orientations in real BaTiO$_3$ crystal. The experimental observations have confirmed such anisotropic electric behaviors and not shown any sign of the saturation polarization [1,8-16,33]. Also the experimental data have exhibited the near linear dependence of the electrostrictive responses on the field, and the polarization jumps in the strain versus field curves. Therefore, our numerical results are in agreement with the experimental observations. The only exception is the strained BaTiO$_3$ thin films in which the saturation polarization was over two times larger than the bulk single crystal for very large compressive strains (~1.7%) [36]. Note that the numerical results shown in Fig. 1 correspond to the static electric field.

However, experimentally the frequency for the applied field can be further reduced to 0.05 Hz [37], and then the quasistatic measurements can be obtained to contrast with the theoretical results.

FIG. 2.(Color online). The electric field induced phase transitions and polarization rotations in BaTiO$_3$ single crystal. (a) The polarization reversal of 180º domains; (b) The polarization rotation and the phase transition for 90º domains; (c) The tetragonal→rhombohedral phase transition.

There are the phase transitions occurring in BaTiO$_3$ single crystal due to the changes in the direction cosines of the polarization accompanying with the polarization rotation. We can determine the phase transitions in terms of the changes in the direction cosines ($\alpha_1,\alpha_2,\alpha_3$). Figure 2 shows the electric field induced phase transitions and polarization rotations in BaTiO$_3$ single crystal. As shown in Fig. 2(a), when the field along the [001] axis decreases, the direction for the polarization of [001]-oriented domains still remains unchanged. Only for $E>E_c$, there is an abrupt polarization reversal. Such polarization reversal may arise from the abrupt atomic displacement of Ti$^{4+}$ and Ba$^{2+}$ ions in BaTiO$_3$ crystal cell with respect to the O$^{2-}$ ions from the [001] direction to the [00$\bar{1}$] diretion, leading to the abrupt change of polarization of the crystal ($P_s$→-$P_s$) [38]. The polarization rotation for 90º domains is more complex. For $E$//[001], the polarization is rotated toward the field direction. In BaTiO$_3$ crystal, $P_1=P_s\alpha_1$, $P_2=P_s\alpha_2$, and $P_3=P_s\alpha_3$. For [100] and [010]-oriented domains, the changes in the order parameters can be given as $P_2^2=0$, $P_1^2>P_3^2$ and $P_1^2=0$, $P_2^2>P_3^2$, respectively, which belong to the monoclinic system. This tetragonal to monoclinic phase transition cannot always go (see Figs. 1(a) and 2(b)). At some critical field, the polarizations are rotated into the field direction in the way similar to the polarization reversal for 180º domains, and the shapes for these domains also can be significantly changed. For $E$//[111], the changes in the order parameters can be indicated by $P_1^2>P_2^2=P_3^2$, $P_1^2=P_2^2<P_3^2$, and $P_1^2=P_3^2<P_2^2$, respectively, which belong to the so-called pseudomonoclinic system. For $E//[110]$, there exist the tetragonal→monoclinic and tetragonal→pseudomonoclinic phase transitions. For $E//[021]$, there are one triclinic phase ($P_1^2>P_2^2>P_3^2$) and two monoclinic phases ($P_1^2=0$, $P_2^2<P_3^2$ and $P_1^2=0$, $P_2^2>P_3^2$). For $E//[123]$, only the tetragonal to triclinic phase transition ($P_1^2>P_2^2>P_3^2$) can occurs. Moreover, when the field is applied along the directions with higher symmetry such as the [112] and [221] axes, there exist two triclinic and one pseudomonoclinic phases. Table I lists various phases existing in BaTiO$_3$ single crystal for the field directions with different symmetries. It has been seen from Table I that the symmetries of the phases are strongly dependent on the symmetries of the field direction.

Table I. Various phases existing in BaTiO$_3$ single crystal for the field directions with different symmetries.

Our results have shown that the changes in the electrostriction for 180º domains are 0 while those for 90º domains are the largest. Accompanying with the 90º polarization reversal, the original [100]-oriented domains can be deformed into [001]-oriented domains in the way that the $c$ axis is rotated by 90º (see in Fig. 2(b)). The largest strain for 90º domains $\varepsilon_{max}$ is dependent on the crystal lattice constants, $a$ and $c$, namely, $\varepsilon_{max}=(c-a)/a$. For the six types of domains in BaTiO$_3$ single crystal, different values for the changes in electrostriction can be generated and then the total changes in electrostriction for the crystal $\Delta\lambda_t$ may enhance or decrease ($\Delta\lambda_t=\Sigma\Delta\lambda_{<100>}V_{<100>}$, $V$ being the fractional occupancies of single domains ). So for large total changes in electrostriction we can make the crystal either be a single domain with large changes in electrostriction, or consist of domains of different orientations with the same electric behaviors.

## IV. CONCLUSIONS

The numerical results have indicated the highly anisotropic polarization rotations and the field-induced-phases with different symmetries in BaTiO$_3$ crystal. In contrast with the conventional LD theory, our theory is physically clear and helpful for understanding the polarization reversal and the electromechanical effect in ferroelectric crystals.

**Acknowledgements**

This work is supported by the Fundamental Research Funds for the Central Universities of China (Grant No. 2012ZM0010).

**Appendix A: Calculation for the electrostriction** [6,39-41]

After the phase transition, the elastic energy $f_{el}$ and the electroelastic energy $f_{ee}$ also are changed dut to the spontaneous deformation.

$$f_{el} = \frac{c_{11}}{2}\left(\varepsilon_{xx}^2 + u_{yy}^2 + u_{zz}^2\right) + c_{12}\left(\varepsilon_{xx}\varepsilon_{yy} + \varepsilon_{yy}\varepsilon_{zz} + \varepsilon_{zz}\varepsilon_{xx}\right) + \frac{c_{44}}{2}\left(\varepsilon_{yz}^2 + \varepsilon_{xz}^2 + \varepsilon_{xy}^2\right)$$

(A1)

$$f_{ee} = -q_{11}P_s^2\left(\varepsilon_{xx}\alpha_1^2 + \varepsilon_{yy}\alpha_2^2 + \varepsilon_{zz}\alpha_3^2\right) - q_{12}P_s^2\left[\varepsilon_{xx}\left(\alpha_2^2 + \alpha_3^2\right) + \varepsilon_{yy}\left(\alpha_3^2 + \alpha_1^2\right) + \varepsilon_{zz}\left(\alpha_1^2 + \alpha_2^2\right)\right]$$
$$-q_{44}P_s^2\left(\varepsilon_{yz}\alpha_2\alpha_3 + \varepsilon_{xz}\alpha_3\alpha_1 + \varepsilon_{xy}\alpha_1\alpha_2\right)$$

(A2)

where $c_{11}$, $c_{12}$, and $c_{44}$ are the elastic constants, $\varepsilon_{ij}(i,j=x,y,z,)$ are the elastic strains, $q_{11}$, $q_{12}$, and $q_{44}$ are the electrostrictive constants, $P_s$ is the saturation polarization, and $\alpha_1$, $\alpha_2$, and $\alpha_3$ are the direction cosines for the polarization. Here we assume that the saturation polarization $P_s$ keeps constant at the temperature $T_0$ below the Curie temperature $T_c$.

The total free energy $f_s$ is

$$f_s = f_{el} + f_{ee} \tag{A3}$$

Under the equilibrium conditions, $\partial f_s/\partial \varepsilon_{ij}=0$, we have

$$\varepsilon_{xx} = k_1\alpha_1^2 + k_2(\alpha_2^2 + \alpha_3^2)$$
$$\varepsilon_{yy} = k_1\alpha_2^2 + k_2(\alpha_3^2 + \alpha_1^2)$$
$$\varepsilon_{zz} = k_1\alpha_3^2 + k_2(\alpha_1^2 + \alpha_2^2) \quad (A4)$$
$$\varepsilon_{yz} = k_3\alpha_2\alpha_3$$
$$\varepsilon_{xz} = k_3\alpha_3\alpha_1$$
$$\varepsilon_{xy} = k_3\alpha_1\alpha_2$$

where

$$k_1 = \frac{(c_{11}+c_{12})q_{11} - 2c_{12}q_{12}}{(c_{11}-c_{12})(c_{11}+2c_{12})}P_s^2$$
$$k_2 = \frac{c_{11}q_{12} - c_{12}q_{11}}{(c_{11}-c_{12})(c_{11}+2c_{12})}P_s^2 \quad (A5)$$
$$k_3 = \frac{q_{44}}{c_{44}}P_s^2$$

When we contrast Eq. (A4) with the LD theory, $k_1 = Q_{11}P_s^2$, $k_2 = Q_{12}P_s^2$, and $k_3 = Q_{44}P_s^2$ ($Q_{11}$, $Q_{12}$, and $Q_{44}$ being the electrostrictive constants in the Gibbs energy function).

The strains along the direction defined by the direction cosines ($\beta_1$, $\beta_2$, $\beta_3$) can be calculated by [6]:

$$\lambda = \varepsilon_{xx}\beta_1^2 + \varepsilon_{yy}\beta_2^2 + \varepsilon_{zz}\beta_3^2 + \varepsilon_{yz}\beta_1\beta_2 + \varepsilon_{xz}\beta_2\beta_3 + \varepsilon_{xy}\beta_3\beta_1$$
$$= k_2 + (k_1 - k_2)(\alpha_1^2\beta_1^2 + \alpha_2^2\beta_2^2 + \alpha_3^2\beta_3^2) + k_3(\alpha_2\alpha_3\beta_2\beta_3 + \alpha_3\alpha_1\beta_3\beta_1 + \alpha_1\alpha_2\beta_1\beta_2) \quad (A6)$$

With the $k_1$, $k_2$, and $k_3$ coefficients replaced with two measured coefficients $\lambda_{100}$ and $\lambda_{111}$ along the [100] and [111] axes, respectively, we have [41]

$$\lambda = \frac{3}{2}\lambda_{100}\left(\alpha_1^2\beta_1^2 + \alpha_2^2\beta_2^2 + \alpha_3^2\beta_3^2 - \frac{1}{3}\right) + 3\lambda_{111}(\alpha_2\alpha_3\beta_2\beta_3 + \alpha_3\alpha_1\beta_3\beta_1 + \alpha_1\alpha_2\beta_1\beta_2) \quad (A7)$$

where $\lambda_{100} = 2/3(Q_{11}-Q_{12})P_s^2$ and $\lambda_{111} = 1/3Q_{44}P_s^2$. We can calculate the electrostrictive responses along different directions in terms of Eq. (A7).

For the uniaxial stress $\sigma$ applied along the direction defined by the direction cosines ($\gamma_1$, $\gamma_2$, $\gamma_3$), $\sigma_{ij} = \sigma\gamma_i\gamma_j$, and the electroelastic energy $f_\sigma$ is

$$f_\sigma = -\sum \sigma_{ij}\varepsilon_{ij}$$
$$= -\frac{3}{2}\lambda_{100}\sigma(\alpha_1^2\gamma_1^2 + \alpha_2^2\gamma_2^2 + \alpha_3^2\gamma_3^2) - 3\lambda_{111}\sigma(\alpha_2\alpha_3\gamma_2\gamma_3 + \alpha_3\alpha_1\gamma_3\gamma_1 + \alpha_1\alpha_2\gamma_1\gamma_2) \quad (A8)$$

The constant term in Eq. (A8) has been neglected.

References

[1] S.-E. Park and T. R. Shrout, J. Appl. Phys. 82, 1804 (1997).

[2] M. K. Durbin, E. W. Jacobs, J. C. Hicks, and S.-E. Park, Appl. Phys. Lett. 74, 2848 (1999).

[3] H. Fu and R. E. Cohen, Nature 403, 281 (2000).

[4] B. Noheda, D. E. Cox, G. Shirane, S.-E. Park, L. E. Cross, and Z. Zhong, Phys. Rev. Lett., 86, 3891 (2001).

[5] D. Damjanovic, IEEE Trans. Ultrason. Ferroelectr. Freq. Control 56, 1574 (2009).

[6] R. C. O'handley, Modern Magnetic Materials: Principles and Applications (John Wiley & Sons, New York, 2000).

[7] M. E. Lines and A. M. Glass, Principles and applications of ferroelectrics and related materials (Clarendon, Oxford, 1979).

[8] C. S. Lynch, Acta Mater. 44, 4137 (1996).

[9] E. A. McLaughlin, T. Liu, and C. S. Lynch, Acta Mater. 52, 3849 (2004).

[10] Q. Wan, C. Chen, and Y. P. Shen, J. Mater. Sci. 41, 2993 (2006).

[11] D. Zhou, M. Kamlah, and D. Munz, J. Eur. Ceram. Soc. 25, 425 (2005).

[12] Q. Wan, C. Chen, and Y. P. Shen, J. Appl. Phys. 98, 024103 (2005).

[13] Z. Feng, D. Lin, H. Luo, S. Li, and D. Fang, J. Appl. Phys. 97, 024103 (2005).

[14] K. G. Webber, Ruzhong Zuo, and C. S. Lynch, Acta Mater. 56, 1219 (2008).

[15] W. Chen and C. S. Lynch, Acta Mater. 46, 5303 (1998).

[16] K. G. Webber, Effect of Domain Wall Motion and Phase Transformations on Nonlinear Hysteretic Constitutive Behavior in Ferroelectric Materials, Ph. D thesis, Georgia Institute of Technology, 2008.

[17] A. E. Clark, in Ferromagnetic materials, Vol. 1, ed. by E. P. Wohlfarth (North Holland, Amsterdam, 1980).

[18] D. C. Jiles, J. Phys. D: Appl. Phys. 27, 1(1994).

[19] J. Atulasimha and A. B. Flatau, Smart Mater. Struct. 20, 043001(2011).

[20] E. C. Stoner and E. P. Wohlfarth, Phil. Trans. Roy. Soc. London A240, 599 (1948), reprinted in IEEE Trans. Magn. 27, 3475 (1991).

[21] W. F. Brown, Jr., Micromagnetics (John Wiley & Sons, New York, 1963), p.29.

[22] G. Bertotti, Hysteresis in Magnetism—for Physicists, Materials Scientists, and Engineers (Academic Press, San Diego, 1998), p.433.

[23] B. Noheda, D. E. Cox, G. Shirane, J. A. Gonzalo, L. E. Cross, and S-E. Park, Appl. Phys. Lett. 74, 2059 (1999).

[24] S.-E. Park, S. Wada, L. E. Cross, and T. R. Shrout, J. Appl. Phys. 86, 2746 (1999).

[25] R. Guo, L. E. Cross, S-E. Park, B. Noheda, D. E. Cox, and G. Shirane, Phys. Rev. Lett. 84, 5423 (2000).

[26] M. Davis, J. Electroceram. 19, 23 (2007).

[27] B. Noheda, D. E. Cox, G. Shirane, R. Guo, B. Jones, and L. E. Cross, Phys. Rev. B. 63, 014103 (2000).

[28] B. Noheda, J. A. Gonzalo, L. E. Cross, R. Guo, S.-E. Park, D. E. Cox, and G. Shirane, Phys. Rev. B. 61, 8687 (2000).

[29] W. Eerenstein, M. Wiora, J. L. Prieto, J. F. Scott, and N. D. Mathur, Nature Mater. 6, 348 (2007).

[30] A. F. Devonshire, Adv. Phys. 3, 85 (1954).

[31] N. A. Pertsev, A. G. Zembilgotov, and A. K. Tagantsev, Phys. Rev. Lett. 80, 1988 (1998).

[32] M. J. Haun, E. Furman, S. J. Jang, H. A. McKinstry, and L. E. Cross, J. Appl. Phys. 62, 3331 (1987).

[33] T. Liu and C.S. Lynch, Acta Mater. 51, 407 (2003).

[34] D. Vanderbilt and M. H. Cohen, Phys. Rev. B. 63, 094108 (2001).

[35] H. Zhang and D. Zeng, J. Appl. Phys. 107, 123918 (2010).

[36] K. J. Choi, M. Biegalski, Y. L. Li, A. Sharan, J. Schubert, R. Uecker, P. Reiche, Y. B. Chen, X. Q. Pan, V. Gopalan, L.-Q. Chen, D. G. Schlom, and C. B. Eom, Science 306,1005 (2004).

[37] A. V. Shil'nikov, A. P. Pozdnyakov, V. N. Nesterov, V. A. Fedorikhin, and R. E. Uzakov, Ferroelectrics 223, 149 (1999).

[38] C. Kittel, Introduction to Solid State Physics (8th edition) (John Wiley & Sons, New Jersey, 2005).

[39] J. Hlinka and P. Márton, Phys. Rev. B. 74, 104104 (2006).

[40] M. Iwata and Y. Ishibashi, in Ferroelectric Thin Films, edited by M. Okuyama

and Y. Ishibashi (Springer-Verlag, Berlin, 2005).

[41] R. E. Newnham, Properties of Materials: Anisotropy, Symmetry, Structure (Oxford University Press, Oxford, 2005).

FIG. 1.(Color online). Dependence of the components of polarization of single domains (a) and electrostrictive responses (b) on the field applied along the [001], [011], and [111] axes. In (a), the inset shows the dependence of the components of the polarization in the field range of -50~50 kV/cm, and the solid arrows indicate the polarization jumps for 180º domains while the dashed arrows indicate the polarization jumps for 90º domains. In (b), the dashed arrows indicate the polarization orientations for single domains at the thermally depolarized state.

FIG. 2.(Color online). The electric field induced phase transitions and polarization rotations in $BaTiO_3$ single crystal. (a) The polarization reversal of 180º domains; (b) The polarization rotation and the phase transition for 90º domains; (c) The tetragonal→rhombohedral phase transition.

| field direction | intermediate phases | | final phases |
|---|---|---|---|
| | order parameters | the symmetry | the symmetry |
| [001] | $P_1^2=0, P_2^2>P_3^2$ | monoclinic | tetragonal |
| | $P_1^2=P_2^2=0, P_3^2=1$ | tetragonal | tetragonal |
| | $P_2^2=0, P_1^2>P_3^2$ | monoclinic | tetragonal |
| [011] | $P_1^2=0, P_2^2<P_3^2$ | monoclinic | tetragonal |
| | $P_2^2=0, P_1^2>P_3^2$ | monoclinic | tetragonal |
| | $P_1^2>P_2^2=P_3^2$ | pseudomonoclinic | tetragonal |
| [021] | $P_1^2=0, P_2^2<P_3^2$ | monoclinic | monoclinic |
| | $P_1^2=0, P_2^2>P_3^2$ | monoclinic | monoclinic |
| | $P_1^2>P_2^2>P_3^2$ | triclinic | monoclinic |
| [111] | $P_1^2>P_2^2=P_3^2$ | pseudomonoclinic | rhombohedral |
| | $P_1^2=P_2^2<P_3^2$ | pseudomonoclinic | rhombohedral |
| | $P_1^2=P_3^2<P_2^2$ | pseudomonoclinic | rhombohedral |
| [112] | $P_1^2=P_2^2<P3$ | pseudomonoclinic | pseudomonoclinic |
| | $P_1^2<P_3^2<P_2^2$ | triclinic | pseudomonoclinic |
| | $P_1^2>P_3^2>P_2^2$ | triclinic | pseudomonoclinic |
| [123] | $P_1^2>P_3^2>P_2^2$ | triclinic | triclinic |
| | $P_1^2<P_2^2<P_3^2$ | triclinic | triclinic |
| | $P_1^2<P_3^2<P_2^2$ | triclinic | triclinic |
| [221] | $P_1^2>P_2^2>P_3^2$ | triclinic | pseudomonoclinic |
| | $P_1^2>P_2^2>P_3^2$ | triclinic | pseudomonoclinic |
| | $P_1^2=P_2^2<P_3^2$ | pseudomonoclinic | pseudomonoclinic |

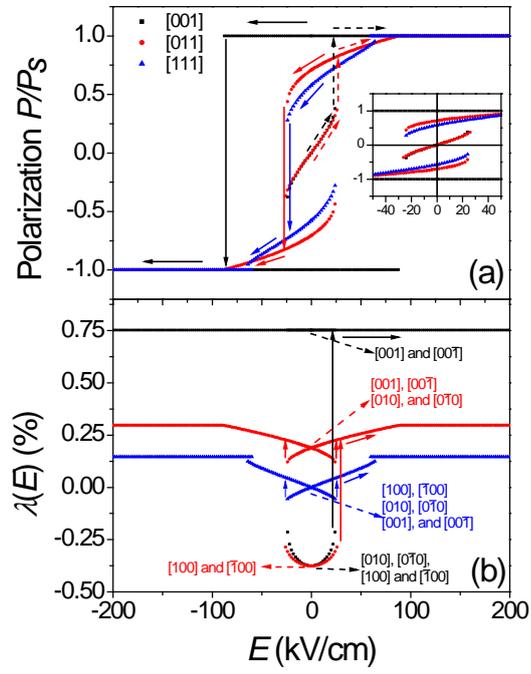

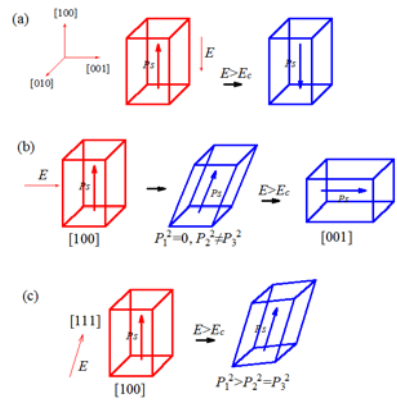